\documentstyle[prl,aps,epsf,multicol]{revtex}
\begin{document}

\draft

\title{Nonequilibrium Electron Distribution in 
Presence of Kondo Impurities}
\author{Georg G\"oppert and Hermann Grabert}
\address{Fakult\"at f\"ur Physik, 
   Albert--Ludwigs--Universit{\"a}t, \\
Hermann--Herder--Stra{\ss}e~3, D--79104 Freiburg, Germany}

\date{\today}
\maketitle
\widetext

\begin{abstract}
We study the energy relaxation of quasiparticles
in voltage biased mesoscopic wires in presence of
magnetic impurities. The renormalization of the 
exchange interaction of Kondo impurities coupled to 
conduction electrons is extended to the case of 
a nonequilibrium electron distribution, which is 
determined self--consistently from a Boltzmann
equation with a collision term due to Kondo impurity
mediated electron--electron scattering.
The approach leads to predictions in quantitative 
agreement with recent experiments by 
Pothier {\it et al.} 
[Phys.\ Rev.\ Lett.\ {\bf 79}, 3490 (1997)].
\end{abstract}


\vspace*{-0.4cm}

\raggedcolumns
\begin{multicols}{2}
\narrowtext

In 1997 the mesoscopic physics community was puzzled
by two experimental findings. On the one hand 
Mohanty, Jariwala, and Webb \cite{MohantyDecPRL97} 
have extracted the electron dephasing time from
weak localization measurements of one--dimensional
gold wires and found larger dephasing rates than 
predicted by the standard theory of Altshuler 
and Aronov (AA) \cite{AltshulerEEint85}. In the 
same year, 
Pothier {\it et al.} \cite{PothierDistrPRL97}
published data on the nonequilibrium electron 
distribution in mesoscopic copper wires in 
presence of an applied voltage. The 
electron--electron scattering rate gained 
from these data was shown to exceed predictions 
based on the AA theory. It
was intuitively clear that these two observations
are very likely due to a common origin, and in 
the last three years a large body
of work has proposed all kinds of mechanisms ranging 
from interaction with two--level systems 
\cite{ZawadowskiDecPRL99}, heating by
radiation \cite{AltshulerPhyE98}, $1/f$ noise 
\cite{ImryDecEPL99} to intrinsic dephasing
at zero temperature \cite{GolubevDecPRL98}.
None of these predictions could give a 
quantitative description of the experiments or 
they were ruled out by subsequent experimental 
studies. Finally, a scenario based on the two 
channel Kondo effect
of symmetrical two level systems has recently been
proposed by Kroha \cite{KrohaASSP00}.
For a rather complete discussion and 
analysis of available experimental data on both
the weak localization and nonequilibrium 
electron distribution measurements in various
metals we refer to the thesis by Pierre 
\cite{PierreThesis00}. 

Already Mohanty {\it et al.} \cite{MohantyDecPRL97}
have demonstrated that  
iron impurities added to the gold wires can
lead to similar effects than those observed
in nominally pure wires. In fact, equating 
the pair breaking rate in superconductors 
containing magnetic impurities \cite{ZittartzPRL71}
with the magnetic contribution to the dephasing
rate in normal metals \cite{HaesendonckPRL87},
one obtains a satisfactory explanation of 
the weak localization data near and above 
the Kondo temperature, where the theory 
is valid \cite{PierreThesis00}.
On the other hand, the effect of Kondo impurities
on electron--electron scattering has been addressed
only very recently.
Kaminski and Glazman (KG) \cite{GlazmanERelax00} 
have determined the two--particle t--matrix
mediated by magnetic impurities for an equilibrium 
system. The renormalization of the exchange coupling
was studied with poor man's scaling using the
applied voltage $V$ as a low--energy cutoff. 
While the theory
can explain the order of magnitude of the 
observed electron--electron scattering rate, 
it does not reproduce the correct voltage 
dependence but leads to more than an order
of magnitude deviations from the data over 
the range of parameters investigated 
experimentally \cite{PothierERelax00}.
In this work we demonstrate that it is essential to 
go beyond poor man's scaling since deviations
from AA predictions are pronounced only 
below the Kondo temperature. Most importantly,
the renormalization flow has to be
determined in presence of a {\it nonequilibrium}
electron distribution calculated self--consistently
from the Boltzmann equation. The collision term then
depends on the distribution function not only through
the occupation probabilities of 
in-- and outgoing electrons
but also via the renormalized interaction kernel.

We start by describing briefly the experimental situation:
A mesoscopic wire of length $L$ with
diffusion constant $D$ is attached to leads biased by a 
voltage $V$. (For details {\it cf.} 
Refs.~\cite{PothierDistrPRL97} and \cite{PothierERelax00}).
The nonequilibrium electron distribution function
$f(\epsilon,x)$ in presence of a steady state
current can be determined from the Boltzmann equation for a
diffusive mesoscopic wire \cite{NagaevPRB95}
\begin{equation}
  \frac{1}{\tau_D} \frac{\partial^2 f(\epsilon,x)}{\partial x^2}
  = 
  I_{\rm coll}
\label{eq:boltzmann}
\end{equation}
with the boundary conditions 
$f(\epsilon,0)=f_F(\epsilon - eV/2)$
and $f(\epsilon,1)=f_F(\epsilon + eV/2)$ imposed by the
leads. Here $x$ is the position within the wire
measured in units of $L$, and $\tau_D=L^2/D$.
In the simple case of vanishing interaction
$I_{\rm coll}=0$ we obtain a double step function
\begin{equation}
 f_0(\epsilon,x)
 =
 (1-x)f_F(\epsilon-eV/2)+x f_F(\epsilon+eV/2).
\label{eq:initialdistr}
\end{equation}
The interaction smears these steps, 
and in the
limit of strong electron--electron interaction leads to
a Fermi function with
an effective temperature of the electrons \cite{PothierZPB97}.
The distribution function $f(\epsilon,x)$ is determined in the
experiment at various locations by tunneling spectroscopy. 

Assuming a local interaction, the collision integral reads
\begin{eqnarray}
\label{eq:icoll}
 I_{\rm coll}(x,\epsilon,\{f\})
&=&
 \int d\omega \int d\epsilon' K(\omega,\epsilon,\epsilon') \\
&&
 \times \big\{
  f(\epsilon)f(\epsilon') [1-f(\epsilon-\omega)][1-f(\epsilon'+\omega)]
 \nonumber  \\
&&
  -
  [1-f(\epsilon)][1-f(\epsilon')] f(\epsilon-\omega)f(\epsilon'+\omega)
 \big\}
 \nonumber
\end{eqnarray}
with an interaction kernel 
$K(\omega,\epsilon,\epsilon')$. Not to overload the notation 
we suppress here and 
in the sequel the spatial dependence of the kernel
and the distribution function.
To determine the kernel  
$K(\omega,\epsilon,\epsilon')$, we follow KG and start from 
the s--d exchange Hamiltonian 
\begin{equation}
 H
 = 
 H_0+H_I
\end{equation}
whereby 
\begin{equation}
 H_0
 = 
 \sum_{k\sigma} \epsilon_k C_{k\sigma}^\dagger C_{k\sigma}
\end{equation}
describes free quasiparticles with one--particle energies
$\epsilon_k$ and creation (annihilation) operators 
$C_{k\sigma}^\dagger$ ($C_{k\sigma}$) of states $k\sigma$. 
We assume that the density of impurities is small enough to
treat the interaction with each impurity independently. Then,
for a single impurity
\begin{equation}
 H_I
 =
 J\sum_{kk'\sigma \sigma'} 
 {\bf S} \cdot {\bf s}_{\sigma \sigma'} 
 C_{k\sigma}^\dagger C_{k'\sigma'},
\end{equation}
where ${\bf S}$ is the impurity spin operator and 
${\bf s}$ the vector of Pauli matrices. 
Further $J$ is the exchange interaction. 

Let us first address the
electron--electron interaction mediated by Kondo
impurities in an equilibrium metal. 
We rewrite the most singular
parts of the interaction in terms of 
single--particle t--matrices and include 
renormalization effects for lower temperatures
by an approach due to Zittartz and M\"uller--Hartmann
\cite{ZittartzZP68,ZittartzSmatrZP68}
based on the Nagaoka equations
\cite{NagaokaPR65}.
This theory, though not able to describe
the zero temperature limit correctly, leads to meaningful 
results at higher temperatures down to 
temperatures well below the Kondo temperature
\cite{Hewson93}.  

\begin{figure}[h]
\begin{center}
\leavevmode
\epsfxsize=0.45 \textwidth
\epsfbox{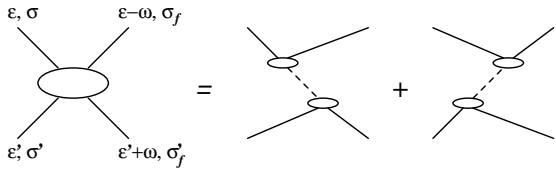}
\vspace*{-.5cm}
\caption{Decomposition of the two--particle t--matrix into two 
single--particle t--matrices.
}
\label{fig:tmatrix}
\end{center}
\end{figure}

For an effective two--particle interaction, the kernel 
$K(\omega,\epsilon,\epsilon')$ is essentially given by
the modulus squared of the on--shell
two--particle t--matrix. Further, one has to sum over all 
final electron spins $\sigma_f, \sigma_f'$ and 
the initial spin $\sigma'$ of the second electron
and average over the initial
spin $\sigma$ of the first electron and the impurity spin 
$S$. For impurities with density $C_{\rm imp}$ the kernel
then takes the form
\begin{equation}
 K(\omega,\epsilon,\epsilon')
 \!=\!
 C_{\rm imp} \rho^3 \frac{\pi}{2\hbar} 
 \sum_{\sigma \sigma_f \sigma' \sigma_f' S} \!
 \langle S| 
  |T_{k \sigma,k' \sigma' \rightarrow 
       k_f \sigma_f,k_f' \sigma_f'} |^2 
 |S \rangle
\label{eq:tmatrfomral}
\end{equation}  
where $\rho$ is the electronic
density of states at the Fermi level. 
The two--particle t--matrix is defined by
\begin{equation}
 T_{k \sigma,k' \sigma' \rightarrow 
       k_f \sigma_f,k_f' \sigma_f'}
 =
 \langle k_f \sigma_f,k_f' \sigma_f' | T 
       | k \sigma,k' \sigma' \rangle 
\label{eq:tmatr}
\end{equation}
and does not depend on the directions of the incoming and 
outgoing electrons. Hence, the wave numbers
$k, k'$ and $k_f, k_f'$ are characterized by the energies
$\epsilon_k=\epsilon$, $\epsilon_{k'}=\epsilon'$, and 
$\epsilon_{k_f}=\epsilon-\omega$, $\epsilon_{k_f'}=\epsilon'+\omega$,
respectively, {\it cf.} Fig.~\ref{fig:tmatrix}.
Unlike an effective potential mediated by
the impurity spin \cite{ZawadowskiPL67}, the t--matrix here
is an operator in impurity spin space. As usual, the
operator $T$ is defined by the series
\begin{equation}
 T(\epsilon)
 =
 H_I \sum_{n=0}^\infty 
 \left(
  \frac{1}{\epsilon-H_0} H_I
 \right)^n.
\label{eq:tmatrop}
\end{equation}

Our interest is in the retarded t--matrix where 
the energy $\epsilon$ is determined by the outgoing 
electrons, i.e.,
$\epsilon=\epsilon_{k_f}+\epsilon_{k_f'}+i\delta$.
Performing perturbation theory up to fifth order 
in the coupling $J$, we get for the inelastic
processes 
\begin{eqnarray}
\label{eq:kernelpt}
 K(\omega,\epsilon,\epsilon') \!
&=& \! 
 \frac{1}{\omega^2} \frac{\pi}{2\hbar}
 \frac{C_{\rm imp}}{\rho} S(S+1) (\rho J)^4  \\
&& \!\!\! 
 \{1 \! + \! \rho J [g(\epsilon) \! + \! g(\epsilon')
     \! + \! g^*(\epsilon-\omega) \! + \! g^*(\epsilon'+\omega)]\}
\nonumber  \\
&&
 + {\rm l.s.} +{\cal O}(\rho J)^6
 \nonumber
\end{eqnarray}
where l.s.\ means less singular terms in $1/\omega$.
To leading order in $J$, this result agrees with the
findings of KG. The less singular terms omitted arise from 
energy denominators that 
include an intermediate electronic energy to be summed over,
leading at most to a logarithmic singularity
for $\omega \rightarrow 0$. Further, 
the auxiliary function $g(\epsilon)$ is given by
\begin{equation}
 g(\epsilon)
 = 
 \int d\epsilon' \left[f(\epsilon')-\frac{1}{2}\right]
 \frac{\tilde{\rho}(\epsilon')}{\epsilon-\epsilon'+i\delta}
\label{eq:pertint}
\end{equation}
where $\tilde{\rho}(\epsilon)$ is the normalized density 
of states of the electrons with $\tilde{\rho}(0)=1$.

To go beyond perturbation theory we consider the 
t--matrix in operator form $(\ref{eq:tmatrop})$.
We are interested in the most divergent terms for 
$\omega \rightarrow 0$ and therefore search for 
energy denominators of the form
$1/(\epsilon-H_0)=\pm 1/\omega$. Since 
$\epsilon=\epsilon_{k_f}+\epsilon_{k_f'}=
 \epsilon_k-\omega+\epsilon_{k'}+\omega = 
 \epsilon_{k}+\epsilon_{k'}$,
the intermediate energy must include both 
$\epsilon_{k_f}$ and $\epsilon_{k'}$ or
$\epsilon_{k_f'}$ and $\epsilon_{k}$, i.e., two 
unperturbed electron lines must lead to this intermediate 
state. Therefore,
in leading order in $1/\omega$ we may decompose the 
two--particle t--matrix into two single--particle 
t--matrices, one acting on the $k$--quasiparticle and 
the other on the $k'$--quasiparticle, 
{\it cf.} Fig.~\ref{fig:tmatrix},
\begin{eqnarray}
\label{eq:tmatrixsing}
T_{k \sigma,k' \sigma' \rightarrow 
       k_f \sigma_f,k_f' \sigma_f'}
&=&
 T_{k \sigma \rightarrow k_f \sigma_f} 
  \frac{1}{-\omega}
 T_{k' \sigma' \rightarrow k_f' \sigma_f'} 
\\
&&
 + \,
 T_{k' \sigma' \rightarrow k_f' \sigma_f'} 
  \frac{1}{\omega}
 T_{k \sigma \rightarrow k_f \sigma_f} 
 + {\rm l.s.} .  \nonumber
\end{eqnarray}
Here, the operator character of the t--matrices in 
impurity spin space plays an essential role since all
commuting terms cancel.

\vspace*{-0.5cm}

To proceed we follow Zittartz \cite{ZittartzSmatrZP68}
and decompose the single--particle t--matrix into a 
non--spin flip and a spin flip amplitude
\begin{equation}
 T_{k \sigma \rightarrow k_f \sigma_f} 
 =
 t_{kk_f}\delta_{\sigma \sigma_f} I_{S} + \tau_{kk_f} 
 {\bf S} \cdot {\bf s}_{\sigma \sigma_f} .
\label{eq:tmatrdec}
\end{equation}
Here, $I_{S}$ means the identity in impurity spin space, and
$t_{kk_f}$ and $\tau_{kk_f}$ are the t--matrices in the 
non--spin flip and spin flip channel, respectively.
Since the identity commutes with other spin matrices, it does
not lead to a leading order contribution in $1/\omega$.
Inserting the two--particle t--matrix 
$(\ref{eq:tmatrixsing})$ with $(\ref{eq:tmatrdec})$
into Eq.~$(\ref{eq:tmatrfomral})$ we get 
\begin{equation}
 K(\omega,\epsilon,\epsilon')
 =
 \frac{1}{\omega^2} \frac{\pi}{2\hbar}
 C_{\rm imp} S(S+1) \rho^3
 |\tau(\epsilon) \tau(\epsilon')|^2  
 + {\rm l.s.}  \,.
\label{eq:kerneltau}
\end{equation}
Comparing this expression with the perturbative result
$(\ref{eq:kernelpt})$, we see that the bare coupling
$J$ is replaced by the renormalized
quantity $\tau_{kk_f}=\tau(\epsilon_{k_f})$. 
This result $(\ref{eq:kerneltau})$ is 
of central importance for the analysis below. 
Note that the expression $(\ref{eq:kerneltau})$ 
displays an explicit $1/\omega^2$ dependence which is 
crucial for the experimentally observed scaling 
behavior \cite{PothierDistrPRL97}.
The kernel for the effective electron--electron
scattering depends on the off--shell spin flip
part of the t--matrix which has a 
different behavior than the on--shell non--spin
flip part of the t--matrix responsible for the
temperature dependence of the resistance. In our approach
the $1/\omega^2$ algebraic factor emerges 
analytically from this representation of 
$K(\omega,\epsilon,\epsilon')$
\cite{kroha2}.

Zittartz \cite{ZittartzSmatrZP68}
has shown that the Nagaoka equations lead to a 
single--particle t--matrix in the spin--flip channel
of the form
\begin{equation}
 \tau(\epsilon)
 = 
 \frac{J}{\phi(\epsilon)}.
\label{eq:tauformal}
\end{equation}
With Hamann's solution \cite{HamannPR67}
of the Nagaoka equations, the denominator reads
\begin{equation}
 \phi(\epsilon)
 =
 [X(\epsilon)^2+S(S+1) (\pi \rho J)^2]^{1/2}.
\label{eq:phihamann}
\end{equation}
Here
\begin{equation}
 X(\epsilon)
 =
 1-S(S+1) (\pi \rho J)^2/4 - \rho J R(\epsilon)
\label{eq:xhamann}
\end{equation}
is temperature dependent via the function
\begin{equation}
 R(\epsilon)
 =
 -\ln\left(\frac{\omega+iT}{i \Lambda} \right)
\label{eq:Reqapprox}
\end{equation}
where $\Lambda$ is the electronic bandwidth. 
$R(\epsilon)$ is an equilibrium approximation
of the auxiliary function $g(\epsilon)$ introduced in  
Eq.~$(\ref{eq:pertint})$. By inserting the solution
$(\ref{eq:tauformal})-(\ref{eq:Reqapprox})$ into
the kernel $(\ref{eq:kerneltau})$ and 
expanding in $J$, we can indeed 
recover the perturbative result $(\ref{eq:kernelpt})$.

Let us now turn to a nonequilibrium situation.
Since in equilibrium and 
nonequilibrium the t--matrix expansions 
differ only in the occupation
probabilities of
intermediate states, we can use the result for the 
equilibrium t--matrix replacing the distribution 
of intermediate 
states by their (unknown) nonequilibrium form.
In the solution 
$(\ref{eq:tauformal})-(\ref{eq:xhamann})$ occupation
probabilities only effect the function $R(\epsilon)$ 
which in a nonequilibrium system 
is no longer of the form $(\ref{eq:Reqapprox})$
but has to be replaced by $g(\epsilon)$.
Of course, then the nonequilibrium 
single--particle t--matrix in the spin flip
channel depends via $g(\epsilon)$ on the 
nonequilibrium distribution function $f(\epsilon)$.
The resulting form of the
collision kernel is fully consistent with 
perturbation theory up to fifth order in $J$ and
it includes correctly the leading 
logarithmic terms.

To proceed, the t--matrix and the distribution function
need to be determined self--consistently. Both quantities 
depend on the position $x$ within the wire. We start 
with the initial nonequilibrium distribution function
$(\ref{eq:initialdistr})$, determine the t--matrix 
from Eqs.~$(\ref{eq:tauformal})-(\ref{eq:xhamann})$,
and the collision kernel from Eq.~$(\ref{eq:kerneltau})$.
An improved distribution function is then obtained 
from the Boltzmann equation $(\ref{eq:boltzmann})$,
which gives rise to an improved kernel. This procedure 
converges after about $20$ iterations.

First, we compared our results with recent experimental
data \cite{PothierERelax00} on the
energy relaxation in Au wires in presence of Fe impurities.
In these experiments all parameters were determined by
independent measurements. For instance, for sample $2$
measured at $T=33$mK the diffusion time 
$\tau_D=1.8$ns. The impurity density 
was determined from the temperature dependence of the 
resistance as $C_{\rm imp}\approx 55$ ppm. 
We set the impurity
spin to $S=1/2$ and the Kondo temperature to 
$T_K =1$K which is a typical value for 
Fe in Au \cite{Hewson93}.
Further, we used a density of states of 
$\rho = 0.25/{\rm site}\, {\rm eV}$ 
\cite{Kittel96}.

\begin{figure}[t]
\begin{center}
\leavevmode
\epsfxsize=0.42 \textwidth
\epsfbox{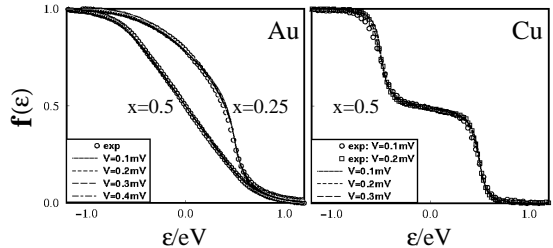}
\vspace*{-.0cm}
\caption{Nonequilibrium distribution function 
for various voltages $V$ and positions $x$
compared with experimental data. Due to
the scaling behavior of $f(\epsilon,V)$, the results
fall on a single line at each position $x$.
}
\label{fig:nonfermi}
\end{center}
\end{figure}

Pothier {\it et al.} \cite{PothierDistrPRL97} 
have noticed that the 
distribution functions $f(\epsilon,V)$ exhibit
scaling behavior 
so that as a function of $\epsilon/eV$ all data 
at a given position $x$ fall on a single line.
In Fig.~\ref{fig:nonfermi} we show our findings
for the distribution function using
the parameters of Pierre {\it et al.}
\cite{PothierERelax00}
for various voltages 
$V=0.1-0.4$meV and two positions $x=0.25, 0.5$. 
They are compared with the experimental data points
\cite{PothierERelax00} and 
we find excellent agreement without adjustable parameters.
We emphasize that the
result is independent of the choice of the bandwidth and is
insensitive to the Kondo temperature as long as $T \ll T_K$.
For other values
of $x$ the curves coincide likewise.

On the other hand, for higher temperatures 
$T>T_K$
where poor man's scaling holds, 
the kernel varies logarithmically with the applied
voltage $V$ \cite{GlazmanERelax00} and 
the distribution function cannot be written
as a function of $\epsilon/eV$ only. 
We have also analyzed the data by 
Pothier {\it et al.} \cite{PothierDistrPRL97}
on Cu wires and find again good agreement  
assuming a density 
$C_{\rm imp}=14$ppm of $S=1/2$ impurities with a
Kondo temperature above $200$mK 
(see Fig.~\ref{fig:nonfermi}). 
Small deviations between the theoretical and experimental
results are likely due to heating 
in this first experiment \cite{PothierPriv01}.

\vspace*{-0.5cm}
\vspace*{-0.4cm}

\begin{figure}[t]
\begin{center}
\leavevmode
\epsfxsize=0.42 \textwidth
\epsfbox{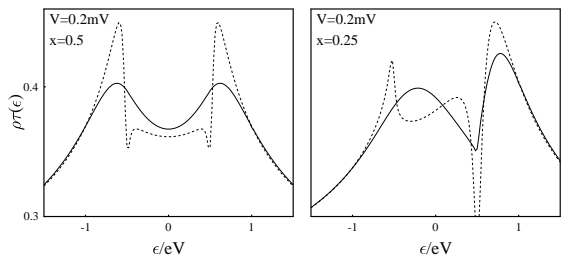}
\vspace*{-.0cm}
\caption{Real part of the t--matrix in the spin--flip channel. 
The solid line shows the self--consistent solution while the
dotted line gives the result using the initial distribution
function $f_0$. 
}
\label{fig:tmatr}
\end{center}
\end{figure}

To see the effect of the nonequilibrium electronic 
distribution on the renormalization of the exchange 
interaction, we show 
in Fig.~\ref{fig:tmatr} for sample $2$ of 
Pierre {\it et al.} \cite{PothierERelax00}
the real part 
of the single--particle t--matrix in the spin flip channel 
$\rho \tau(E)$ for $V=0.2$meV at 
$x=0.5$ (left panel) and 
$x=0.25$ (right panel). The solid line gives the 
self--consistent solution obtained from the iteration explained 
above, while the dotted line depicts the result obtained 
for the distribution $(\ref{eq:initialdistr})$
in absence of energy relaxation.
We see that the coupling changes significantly
with the distribution function. Using only the initial
t--matrix does not suffice to explain the
experimental data. Comparing t--matrices for
various voltages, we find weak dependence
between the Fermi points $\epsilon=\pm eV/2$. This
gives rise to the scaling behavior of the distribution
function, while poor man's scaling \cite{GlazmanERelax00}
implies a significant voltage dependence of the scaled
data in conflict with experiments \cite{PothierERelax00}. 

\vspace*{-0.5cm}
\vspace*{-0.5cm}

To complete our discussion we should have included 
spin dynamics which cuts off the $1/\omega^2$ 
divergence of $K(\omega,\epsilon,\epsilon')$ for small 
frequencies \cite{GlazmanERelax00}.
This is crucial for effects that depend strongly 
on the low frequency limit, such as dephasing. 
However, the distribution function 
$f(\epsilon)$ determined from the Boltzmann equation 
$(\ref{eq:boltzmann})$ is almost 
insensitive to this low frequency cutoff since the
collision integral remains finite even in absence 
of a cutoff. Hence, for the problem 
considered here, one can disregard the impurity spin 
relaxation.

In summary we have determined the 
effective electron--electron collision kernel mediated
by magnetic impurities for nonequilibrium wires
at temperatures below the Kondo temperature. We 
found excellent agreement with recent experimental 
findings \cite{PothierDistrPRL97,PothierERelax00}. 
In particular, we
have demonstrated that the distribution function
displays scaling only in the regime below the
Kondo temperature.

The authors would like to thank the authors of 
Ref.~\cite{PothierERelax00} for valuable discussions.
Financial support was provided by the Deutsche
Forschungsgemeinschaft (DFG) and the Deutscher Akademischer 
Austauschdienst (DAAD).

\vspace*{-0.5cm}

%

\begin{thebibliography}{10}
\bibitem{kroha2} 
 This is one of the main differences between this
 approach and the NCA techniques used in Ref \cite{KrohaASSP00}
 where algebraic behavior only arises from a summation of
 an infinite series of logarithmic corrections.
\end{thebibliography}
%

%
\end{multicols}
\end{document}